\documentclass{moriond}

\def\be{\begin{equation}}
\def\ee{\end{equation}}
\def\bea{\begin{eqnarray}}
\def\eea{\end{eqnarray}}

\newcommand{\Photo}{}

\begin{document}
\vspace*{4cm}
\title{Relic neutrino decay: a solution to the excess radio background}

\author{ Rishav Roshan }

\address{School of Physics and Astronomy, University of Southampton,\\ Southampton, SO17 1BJ, U.K.}

\maketitle

\abstract{
The detection of excess radio background detected by ARCADE 2 suggests the presence of new physics as there exists no clear astrophysical solution. We find that the radiative decay of a relic neutrino into a sterile neutrino, assumed to be quasi-degenerate, provides a very good fit to ARCADE 2 data. The solution also predicts a stronger 21 cm absorption global signal than the predicted one from the $\Lambda$CDM model, with a contrast brightness temperature. The solution obtained is in mild tension with a much stronger signal claimed by the EDGES collaboration.}

\section{Introduction}
The measurement of extragalactic radio excess in addition to the cosmic microwave background (CMB) approximately in the frequency range  3 to 8 GHz by Absolute Radiometer for Cosmology, Astrophysics and Diffuse Emission (ARCADE 2)\cite{Fixsen:2009xn} has dazzled the radioastronomy community. Since the excess radio background (ERB) cannot be explained by known sources, it motivates us to look for physics beyond the Standard Model (BSM). This observed ERB in the present universe  ($z=0$) has to be attributed to redshifted radiation produced in the early universe (high redshift). This would produce a sizeable deviation of the cosmological 21 cm global absorption signal from the standard prediction ($T_{21}^{\Lambda\rm CDM} = -200\,{\rm mK}$). Such a deviation was indeed claimed by the Experiment to Detect the Reionization Step (EDGES) collaboration recently \cite{Bowman:2018yin}. EDGES observes an absorption signal ($T_{21}^{\rm EDGES} = -500^{+200}_{-500}\,{\rm mK}$ ($99\%\,{\rm C.L.}$))at $z_E\simeq 17$ that is approximately twice as strong as the standard one. So far the claim has not yet been confirmed by any other experiments.

In this proceeding which is based on the work\cite{Dev:2023wel}, we focus our attention on the radiative decay of lightest relic neutrino as a solution of the ERB observed in the present universe. The solution is interesting because it depends only on two parameters: the lifetime of neutrinos ($\tau_1$) and the mass difference ($\Delta m_1$) between the relic neutrinos and invisible particles into which they decay. On top of this relic neutrino background is fixed as one can assume it to be the standard one predicted by the $\Lambda$CDM model. On the other hand, since the neutrinos are fermion, they do not cluster much and hence predict a very smooth ERB as supported by the observation. We show that the solution provides a very good fit for ARCADE 2 observation and also predicts a stronger 21 cm absorption global signal than the $\Lambda$CDM model, but our prediction is milder than the absorption signal observed by the EDGES collaboration.

\section{Radiative decay of relic neutrinos}
We consider the production of non-thermal photons from the decay of the lightest relic neutrino (with mass $m_1$) $i.e~\nu_1\to\nu_s\gamma$ where $\nu_s$ stands for a sterile neutrino with mass $m_s$. We also introduce $\Delta m_1=m_1-m_s<<m_1$. In the case of ARCADE 2, since the measurements are made at $z=0$, the effective ERB temperature from relic neutrino decays can be approximated as
\begin{equation}\label{Tgammanth}
T_{\gamma_{\rm nth}}(E,0) \simeq  {6\,\zeta(3)\over 11 \, \sqrt{\Omega_{{\rm M}0}}}\,
{T_{0}^3 \over E^{1 / 2}\, \Delta m_1^{3 / 2} } \, {t_0 \over \tau_1} 
\,\left(1 +{a_{\rm D}^3 \over a_{\rm eq}^3} \right)^{-{1 \over 2}} \,  
\end{equation}
as $E<<T_\gamma$ with $E$ being the energy of non-thermal photons. Here, $\Omega_{{\rm M}0}\simeq0.3111$ denotes matter energy density parameters at present time $t_0$, $a_{\rm D}$ represents the scale factor at the time of decay, and $a_{\rm eq}\simeq 0.77$. Note that here we always have $E\leq\Delta m_1$ which necessarily implies the {\em existence of an endpoint} $E_{\rm max} =\Delta m_1$. This is a very clear prediction of the model that will be tested by the experiments aiming at detecting deviations from CMB such as PIXIE \cite{Kogut:2011xw} and TMS \cite{Alonso-Arias:2021quq}. We also place an upper bound $\Delta m_1 < 2.5 \times 10^{-4}\,{\rm eV}$,  corresponding to frequencies $\nu < 60 \,{\rm GHz}$, since at higher frequencies there are very stringent constraints from the FIRAS \cite{2002ApJFIRAS}. After doing the $\chi^2$ analysis, we find a perfect fit for ARCADE 2 data with $\tau_1=1.46\times 10^{14}$s and $\Delta m_1=4.0\times10^{-5}$ eV.

Another \emph{smoking gun test} for our relic neutrino decay hypothesis is the prediction of non-thermal photons at higher redshifts, can be provided by the observation of the 21 cm cosmological global absorption signal. Such an absorption signal was also observed by the EDGES collaboration. In the case of EDGES the detection occurs at $z \sim z_E$ and it is seen today as a (redshifted) absorption feature. The 21 cm signal can be expressed in terms of brightness temperature, 
\begin{equation}\label{T21}
T_{21}(z) \simeq  23\,{\rm mK} \, (1+\delta_{\rm B})\, x_{H_I}(z) \,\left({\Omega_{{\rm B}0} h^2 \over 0.02}\right)\,
\left[\left({0.15 \over \Omega_{{\rm M}0}h^2}\right)\,
\left({1+z \over 10} \right)  \right]^{1/2}  \,\left[
1 - {T_\gamma(z) \over T_{\rm S}(z)} \right]  \,  ,
\end{equation}
where $\delta_{\rm B}$ is the fractional baryon overdensity, $x_{H_I}(z)$ is the neutral hydrogen fraction, $\Omega_{{\rm B}0}$
is the baryon and matter energy density parameters. In present case the non-thermal photon temperature at $z_E$ can be obtained by 
\begin{equation}\label{TgammanthEDGES}
T_{\gamma_{\rm nth}}(E_{21},z_E) \simeq  {6\,\zeta(3)\over 11 \, \sqrt{\Omega_{{\rm M}0}}}\,
{T_{0}^3\,(1 +z_E)^{3/2}\over E_{21}^{1 / 2}\, \Delta m_1^{3 / 2} } \, {t_0 \over \tau_1}  \,  ,
\end{equation} 

\noindent Substituting the best fit values of $\Delta m_1$ and $\tau_1$ mentioned in the last paragraph into Eq.~(\ref{TgammanthEDGES}), finding $T_{\gamma_{\rm nth}}(E_{21},z_E)$, and then further substituting it in Eq.~(\ref{T21}) one obtains:
\begin{equation}\label{T21predicted}
T_{21}(\bar{z}_E) = -238^{+21}_{-20}\,{\rm mK} \;\;\; (99\%\,{\rm C.L.}) . 
\end{equation}  
As can be seen, our model predicts a stronger absorption signal compared to $\Lambda$CDM but milder in comparison to that of the EDGES.

\section{Final Remarks}
We considered a case of the lightest neutrino decaying into sterile neutrinos and non-thermal photons and find a very good fit to the ARCADE 2 data with best-fit values $\tau_i = 1.46  \times 10^{21}\,{\rm s}$ suggesting the \emph{ existence of an endpoint} that can be tested can be tested in the next years by TMS in the 10-20 GHz frequency range. Finally, our solution also predicts a stronger 21 cm absorption global signal with $T_{21} = -238^{+21}_{-20}\,{\rm mK}$ ($99\%$ C.L.) at redshift $z_E$. 
\section{Acknowledgments}
RR acknowledges financial support from the STFC Consolidated Grant ST/T000775/1 which has made his participation possible in the 58th Rencontres de Moriond.

\end{document}